\documentclass[11pt,a4paper,twoside,fleqn]{article}
\usepackage{color,graphicx,natbib,lscape,here,geometry,setspace,multirow,enumitem}
\usepackage[dvipsnames]{xcolor}
\usepackage{graphicx}
\usepackage{authblk}
\usepackage{silence}
\usepackage{rotating}
\usepackage{multirow}

\usepackage{booktabs}
\usepackage{diagbox}
\usepackage{color}
\usepackage{setspace}
\usepackage{algorithm2e}
\usepackage{enumitem}
\usepackage{tikz}

\definecolor{nblue}{HTML}{000660}
\usepackage[colorlinks=true,urlcolor=nblue,linkcolor=nblue,citecolor=nblue]{hyperref}
\usepackage{bigstrut}

\usepackage{tabularx}
\usepackage{threeparttable}
\usepackage{dcolumn}
\newcolumntype{d}[1]{D{.}{.}{#1}}

\usepackage{array}
\newcolumntype{C}[1]{>{\centering\arraybackslash}p{#1}}

\usepackage{etoolbox} 
\makeatletter
\patchcmd{\BR@backref}{\newblock}{\newblock[}{}{}
\patchcmd{\BR@backref}{\par}{]\par}{}{}
\makeatother

\usepackage{pdflscape}
\usepackage{afterpage}

\usepackage{longtable}
\usepackage{array}

\usepackage{comment}
\usepackage{mathtools}

\usepackage[hang,flushmargin]{footmisc} 

\usepackage[hang]{footmisc}
\usepackage[british]{babel}

\pdfoutput=1
\usepackage{float}
\restylefloat{table}

\usepackage[title,titletoc]{appendix}
\makeatletter

\renewenvironment{appendices}{%
    \begin{oldappendices}%
    \renewcommand{\thefigure}{\ifnum \c@section>\z@ \thesection.\fi\@arabic\c@figure}%
    \@addtoreset{figure}{section}%
    \renewcommand{\thetable}{\ifnum \c@section>\z@ \thesection.\fi\@arabic\c@table}%
    \@addtoreset{table}{section}}{%
    \end{oldappendices}%
}\makeatother

\usepackage{titlesec} 
\titleformat{\section}[block]{\large}{\thesection. }{0em}{\MakeUppercase} 
\titleformat{\subsection}[block]{\large}{\thesubsection. }{0em}{\itshape} 
\titleformat{\subsubsection}[block]{\large}{}{0em}{\itshape} 

\let\natbibcitet\citet
\renewcommand\citet{\bibpunct{(}{)}{,}{a}{,}{,}\natbibcitet}
\let\natbibcitep\citep
\renewcommand\citep{\bibpunct{(}{)}{;}{a}{,}{;}\natbibcitep}
\newcommand{\bi}{\begin{itemize}}
\newcommand{\ei}{\end{itemize}}
\newcommand{\be}{\begin{equation}}
\newcommand{\ee}{\end{equation}}
\defcitealias{ieo14}{IEO, 2014}

\long\def\symbolfootnote[#1]#2{\begingroup%
\def\thefootnote{\fnsymbol{footnote}}\footnote[#1]{#2}\endgroup}

\makeatletter
\def\ubar#1{\underline{\sbox\tw@{$#1$}\dp\tw@\z@\box\tw@}}
\def\obar#1{\overline{\sbox\tw@{$#1$}\dp\tw@\z@\box\tw@}}
\makeatother

\usepackage{bm}
\usepackage{caption}
\usepackage{subcaption}

\makeatletter\let\p@subfigure\thefigure\makeatother

\floatstyle{plaintop}
\restylefloat{table}

\usepackage{fancyhdr} 
\usepackage{cleveref}
\crefname{chapter}{Chapter}{Chapters}
\crefname{section}{Section}{Sections}
\crefname{subsection}{Section}{Sections}
\crefname{subsubsection}{Section}{Sections}
\crefname{figure}{Figure}{Figures}
\crefname{table}{Table}{Tables}
\crefname{equation}{Equation}{Equations}
\crefname{appendix}{Appendix}{Appendices}
\crefname{appendices}{Appendix}{Appendices}

\crefname{appsec}{Appendix}{Appendices}

\usepackage{catoptions}
\makeatletter

\def\Autoref#1{%
  \begingroup
  \edef\reserved@a{\cpttrimspaces{#1}}%
  \ifcsndefTF{r@#1}{%
    \xaftercsname{\expandafter\testreftype\@fourthoffive}
      {r@\reserved@a}.\\{#1}%
  }{%
    \ref{#1}%
  }%
  \endgroup
}
\def\testreftype#1.#2\\#3{%
  \ifcsndefTF{#1autorefname}{%
    \def\reserved@a##1##2\@nil{%
      \uppercase{\def\ref@name{##1}}%
      \csn@edef{#1autorefname}{\ref@name##2}%
      \autoref{#3}%
    }%
    \reserved@a#1\@nil
  }{%
    \autoref{#3}%
  }%
}
\makeatother

\title{\LARGE{\textbf{Dynamic shrinkage in time-varying parameter stochastic volatility in mean models}}}
\author{\large{
\uppercase{Florian Huber} and \uppercase{Michael Pfarrhofer}}\thanks{
\noindent \textit{Corresponding author}: Michael Pfarrhofer. Salzburg Centre of European Union Studies, University of Salzburg. \textit{Address}: M\"{o}nchsberg 2a, 5020 Salzburg, Austria. \textit{Email}: \href{mailto:michael.pfarrhofer@sbg.ac.at}{michael.pfarrhofer@sbg.ac.at}. \textit{Phone}: +43 662 8044 3772. The authors gratefully acknowledge financial support from the Austrian Science Fund (FWF, grant no. ZK 35).}
\\\vspace*{-0.5em}
\textit{University of Salzburg}}
\date{}


\def\equationautorefname~#1\null{%
  Eq.~(#1)\null
}
\def\equationautorefname~#1\null{
Eq.~(#1)\null
}

\usepackage[utf8, latin1]{inputenc}                 

\begin{document}
\maketitle\thispagestyle{empty}\normalsize\vspace*{-2em}\small

\begin{center}
\begin{minipage}{0.8\textwidth}
\noindent\small Successful forecasting models strike a balance between parsimony and flexibility. This is often achieved by employing suitable shrinkage priors that penalize model complexity but also reward model fit. In this note, we modify the stochastic volatility in mean (SVM) model proposed in \cite{chan2017stochastic} by introducing state-of-the-art shrinkage techniques that allow for time-variation in the degree of shrinkage. Using a real-time inflation forecast exercise, we show that employing more flexible prior distributions on several key parameters slightly improves forecast performance for the United States (US), the United Kingdom (UK) and the Euro Area (EA). Comparing in-sample results reveals that our proposed model yields qualitatively similar insights to the original version of the model.
\end{minipage}
\end{center}

\onehalfspacing\normalsize\renewcommand{\thepage}{\arabic{page}}

\section{Introduction}\label{sec:introduction}
Forecasting in macroeconomics and finance requires flexible models that are capable of capturing salient features of the data such as structural breaks in the regression coefficients and/or heteroscedastic measurement errors. Time-variation in the shocks is often introduced through stochastic volatility (SV) models that imply a smoothly evolving error variance over time. Such models typically rule out that the level of the volatility directly affects the conditional mean of the predictive regression. This assumption is relaxed in \citet{koopman2002stochastic} and \citet{chan2017stochastic} by assuming that the volatilities enter the conditional mean equation and thus exert a direct effect on the quantity of interest.

In this note, we reconsider the model proposed in \cite{chan2017stochastic} and replicate the main findings both in a narrow and wide sense. The original specification is a time-varying parameter (TVP) model with SV that allows for feedback effects between the level of volatility and the endogenous variable. As opposed to most of the existing literature, this model assumes that this relationship is time varying. Estimation and inference is carried out in a Bayesian framework, implying that prior distributions are specified on all coefficients of the model. These priors are often set to be weakly informative.

One key contribution of this note is to introduce shrinkage via state-of-the-art dynamic shrinkage priors that allow for capturing situations where coefficients are time-varying over certain periods in time while they remain constant in others.\footnote{A similar exercise using a mixture innovation model is provided in \citet{hou2020time}.} These priors are based on a recent paper, \cite{kowal2019dynamic}, that proposes introducing a dynamic shrinkage process that is time-varying and follows an AR(1) model with Z-distributed shocks. Proper specification of the hyperparameters of this error distribution yields a dynamic Horseshoe (DHS) prior that possesses excellent shrinkage properties. Other specifications we propose also introduce shrinkage but assume the shrinkage coefficients to be independent over time (static horseshoe prior, SHS) or time-invariant, such as a standard horseshoe (HS) prior that exploits the non-centered parameterization of the state space model \citep[see][]{FRUHWIRTHSCHNATTER201085}.

The second contribution deals with replicating the main findings of \cite{chan2017stochastic} using updated real-time inflation data. Instead of considering the original three countries (the US, the UK and Germany), we replace Germany with the EA and investigate whether the main findings also hold for this dataset. Using more flexible shrinkage priors generally yields similar in-sample findings for the US and the UK. For the EA, we find only minor evidence of a link between inflation and inflation volatility. This finding relates to \cite{jarocinski2018inflation}, who observe limited evidence in favor of SV for inflation derived from the harmonized index of consumer prices (HICP). When it comes to forecasting we find that shrinkage sometimes improves predictive accuracy. In cases where predictive accuracy is below the no-shrinkage specification, these differences are often very small.

In the remainder of the note we proceed as follows. The next section summarizes the model and motivates our shrinkage priors. Section \ref{sec:application} replicates the main findings of \cite{chan2017stochastic} using the proposed model and carries out a real-time forecasting exercise to show that using shrinkage often further improves upon the already excellent predictive performance of the original model. Finally, the last section briefly summarizes and concludes the note.

\section{Econometric Framework}\label{sec:econometrics}
\subsection{The Time-varying Parameter Stochastic Volatility in Mean Model}
The time-varying parameter stochastic volatility in mean (TVP-SVM) model is given by:
\begin{align}
y_t &= \tau_t + \bm{\beta}_t'\bm{z}_t + \gamma_t e^{h_t} + \epsilon_t, \quad \epsilon_t \sim \mathcal{N}\left(0, e^{h_t}\right),\label{eq: obs}\\
h_t &= \mu_h + \phi_h(h_{t-1} - \mu_h) + \delta y_{t-1} + \nu_{t}, \quad \nu_t \sim \mathcal{N}(0, \sigma^2),
\end{align}
where $y_t$ is a scalar time series, $\tau_t$ denotes a stochastic trend term, $\bm \beta_t$ is a $K$-dimensional vector of dynamic regression coefficients while $\gamma_t$ is a coefficient that measures the (potentially) time-varying relationship between $y_t$ and the shock volatility $e^{h_t}$.  The column vector $\bm{z}_t$ may contain lags of the dependent variable, additional predictors and/or latent factors capturing high-dimensional information. The log-volatility $h_t$ follows an AR(1) process with unconditional mean $\mu_h$, persistence parameter $\phi_h$, and error variance $\sigma^2$. $h_t$, moreover, depends on the lag of $y_t$ through a time-invariant parameter $\delta$.

Let $\bm{x}_t = (1,\bm{z}_t',e^{h_t})'$ and $\bm{\theta}_t = (\tau_t,\bm{\beta}_t',\gamma_t)'$ of size $k \times1$ (with $k=2+K$), then \autoref{eq: obs} can be written in regression form:
\begin{equation}
 y_t = \bm{\theta}_t'\bm{x}_t + \epsilon_t, \quad \epsilon_t \sim \mathcal{N}\left(0, e^{h_t}\right).
\end{equation}
Furthermore, we assume that $\bm \theta_t$ evolves according to a random walk:
\begin{equation}
\bm{\theta}_t = \bm{\theta}_{t-1} + \bm{e}_t, \quad \bm{e}_t\sim\mathcal{N}(\bm{0},\bm{\Omega}),
\end{equation}
with Gaussian errors and diagonal covariance matrix $\bm{\Omega}=\text{diag}(\omega_1,\hdots,\omega_k)$. 

\subsection{Imposing Shrinkage in TVP Models}
The model outlined in the previous sub-section is quite flexible and allows for a direct relationship between the error volatilities and $y_t$. And this relationship might be subject to parameter instability. Allowing for TVPs in all coefficients could, however, lead to overfitting and this often leads to decreases in predictive accuracy. \cite{chan2017stochastic} uses weakly informative priors on key parameters and finds them to yield good forecasting results. 

Here, we aim to improve upon this finding by introducing three additional priors that allow us to flexibly select restrictions in the empirical model and thus achieve parsimony. The priors we consider in this study are given by:
\begin{enumerate}[leftmargin=1.5em,itemsep=0em,label=(\arabic*)]
	\item A weakly informative prior on the coefficients and state innovation variances similar as in \citet{chan2017stochastic}. We use independent weakly informative inverse gamma priors on the innovation variances of the state equation $\omega_j$. We subsequently label this prior ``None,'' reflecting the notion that almost no shrinkage is imposed.
	\item A hierarchical global local prior on the constant part and innovation variances of the model. We achieve this by rewriting the model in the non-centered parameterization of \citet{FRUHWIRTHSCHNATTER201085}:
	\begin{align}
	y_t &= \bm{\theta}_0'\bm{x}_t + \bm{\tilde\theta}_t'\sqrt{\bm{\Omega}}\bm{x}_t + \epsilon_t,\\
	\bm{\tilde\theta}_t &= \bm{\tilde\theta}_{t-1} + \bm{\eta}_t, \quad \bm{\eta}_t\sim\mathcal{N}(\bm{0}_k,\bm{I}_k)
	\end{align}
	with $\sqrt{\bm{\Omega}} = \text{diag}(\sqrt{\omega_1},\hdots,\sqrt{\omega_k})$, the $j$th element of $\tilde\theta_{jt} = (\theta_{jt}-\theta_{j0})/\sqrt{\omega_j}$ and $\bm{\tilde\theta}_0=\bm{0}_k$. We collect the constant parameters and the state innovation variances in a $2k\times1$-vector $\bm{\alpha} = (\bm{\theta}_0',\sqrt{\omega_1},\hdots,\sqrt{\omega_k})'$ and index its $i$th element for $i=1,\hdots,2k$ by $\alpha_i$. Any shrinkage prior on these coefficients may be used, and we rely on the popular horseshoe prior (labeled ``HS'' in the empirical application) of \citet{carvalho2010horseshoe} in its auxiliary representation \citep{makalic2015simple}:
	\begin{equation}
	\alpha_i \sim \mathcal{N}(0,\phi_i\lambda), \quad \phi_i\sim\mathcal{G}^{-1}(1/2,1/v_i), \quad \lambda\sim\mathcal{G}^{-1}(1/2,1/w),
	\end{equation}
	with $v_i\sim\mathcal{G}^{-1}(1/2,1)$ for $i=1,\hdots,2k$ and $w\sim\mathcal{G}^{-1}(1/2,1)$.
	\item A static variant of the horseshoe prior (labeled ``SHS'') imposing shrinkage using the centered parameterization of the state equation with time-varying variances:
  \begin{equation}
  \bm{\theta}_t = \bm{\theta}_{t-1} + \bm{e}_t, \quad \bm{e}_{t}\sim\mathcal{N}(\bm{0},\bm{\Omega}_{t}).
  \end{equation}
  We denote the $j$th diagonal element of $\bm{\Omega}_t$ by $\omega_{jt}=\lambda_j\phi_{jt}$ and assume inverse Gamma distributions as priors for the global and local shrinkage parameters
  \begin{equation}
  \phi_{jt}\sim\mathcal{G}^{-1}(1/2,1/v_{jt}), \quad \lambda_j\sim\mathcal{G}^{-1}(1/2,1/w_j).
  \end{equation} 
  Following \citet{makalic2015simple}, auxiliary variables $v_{jt}\sim\mathcal{G}^{-1}(1/2,1)$ and $w_j\sim\mathcal{G}^{-1}(1/2,1)$ for $j=1,\hdots,k$ are used for establishing the horseshoe prior. Here, $\lambda_j$ governs the overall amount of time variation for the coefficient of the $j$th regressor, while $\varphi_{jt}$ allows for predictor and time specific shrinkage.
	\item A dynamic horseshoe prior (labeled ``DHS'') as in \citet{kowal2019dynamic}. Again using the centered parameterization of the state equation with time-varying state innovation variances in $\bm{\Omega}_t$ with $j$th element $\omega_{jt}=\lambda_0\lambda_j\phi_{jt}$. To achieve a log-scale representation of the global local prior, define $\psi_{jt} = \log(\lambda_0\lambda_j\phi_{jt})$  and assume
	\begin{equation}
	\psi_{jt} = \mu_{\psi j} + \varphi_j(\psi_{t-1} - \mu_{\psi j}) + \nu_{jt}, \quad \nu_{jt}\sim\mathcal{Z}(a,b,0,1),
	\end{equation}
	with $\mathcal{Z}$ denoting the Z-distribution, where setting $a=b=1/2$ yields the dynamic horseshoe prior \citep[for details on related prior choices, see][]{kowal2019dynamic}. Here $\lambda_0$ is a global, $\lambda_j$ are predictor specific, and $\phi_{jt}$ are predictor and time-specific shrinkage parameters that follow a joint autoregressive law of motion.
\end{enumerate}

We use standard Markov chain Monte Carlo (MCMC) methods such as Gibbs sampling augmented by a forward filtering backward sampling (FFBS) algorithm for the TVPs \citep{doi:10.1093/biomet/81.3.541,doi:10.1111/j.1467-9892.1994.tb00184.x}. For the log-volatilities related to the dynamic shrinkage prior, the procedure outlined in \citet{kowal2019dynamic} employing a mixture representation of the Z-distribution using P\'olya-gamma random variables is applicable. The SV processes are simulated by adapted independent Metropolis-Hastings updates as proposed by \citet{kim1998stochastic}, with a prior setup as in \citet{KASTNER2014408}. Our algorithm is implemented in \texttt{R}, providing further robustness to the findings from the \texttt{MATLAB} implementation in the original contribution.

\section{Inflation Modeling}\label{sec:application}
In this study we take a real time perspective to modeling inflation for the US, UK, and the EA. Vintage data available at specific times in the past is obtained from the webpages of the Federal Reserve Bank of St. Louis (\href{https://research.stlouisfed.org/econ/mccracken/fred-databases/}{fred.stlouisfed.org}), Bank of England (\href{https://www.bankofengland.co.uk/statistics/gdp-real-time-database}{bankofengland.co.uk}), and the European Central Bank (\href{https://sdw.ecb.europa.eu/browseExplanation.do?node=9689716}{sdw.ecb.europa.eu}).

Price indices $p_t$ taken from the respective databases are seasonally adjusted and on quarterly frequency (taking the average over the respective months if on higher frequency originally). For the US, we use the consumer price index (CPIAUCSL), the gross domestic product deflator at market prices (PGDPDEF) for the UK, and the harmonized index of consumer prices (HICP) for the EA. Historical vintage data for the US, UK and EA start in 1998, 1990 and 2001, resulting in differently sized natural holdout samples with a total available time period ranging from 1959:Q1 to 2019:Q1 (US), 1970:Q1 to 2016:Q3 (UK), and 1996:Q1 to 2019:Q1 (EA), respectively.

We model inflation, defined as $\pi_t = 400\log(p_t/p_{t-1})$, with an unobserved component model augmented with stochastic volatility in the mean (UC-SVM):
\begin{align*}
\pi_t &= \tau_t + \gamma_t e^{h_t} + \epsilon_t, \quad \epsilon_t \sim \mathcal{N}(0, e^{h_t})\\
h_t &= \mu_h + \phi_h(h_{t-1} - \mu_h) + \delta \pi_{t-1} + \nu_{t}, \quad \nu_t \sim \mathcal{N}(0, \sigma^2)\\
\tau_t &= \tau_{t-1} + e_t, \quad e_t \sim \mathcal{N}(0, \omega),
\end{align*}
which is a special case of Eq. (\ref{eq: obs}) with $\bm \beta_t = \bm 0$ for all $t$. This model has been used by \cite{chan2017stochastic} to forecast inflation. If $\gamma_t = 0$, we obtain the UC-SV model proposed in \cite{stock2007has}.  If the prior on the state innovation variance $\omega$ is specified too loose, the model might be prone to overfitting and this would be deleterious for predictive accuracy. Hence, in this empirical application we assess whether using shrinkage priors improves the predictive fit of the model. But before we turn to analyzing predictions, we focus on key in-sample results.

\subsection{In-sample results}
\begin{figure}[!htbp]
\begin{center}
\begin{subfigure}[b]{\textwidth}
\caption{United States}
\includegraphics[width=\textwidth]{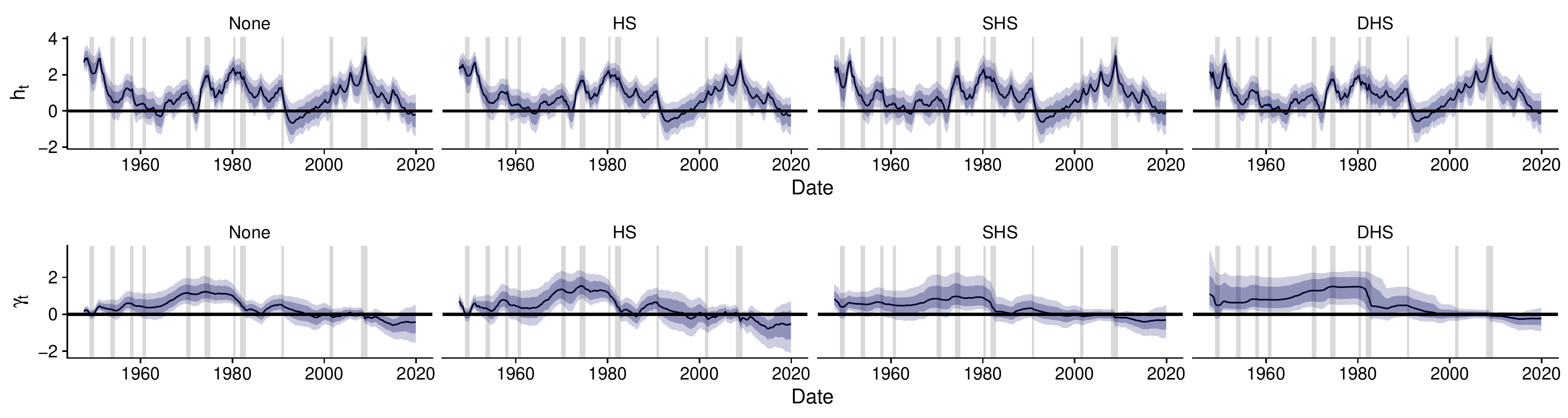}
\end{subfigure}
\begin{subfigure}[b]{\textwidth}
\vspace*{1em}\caption{United Kingdom}
\includegraphics[width=\textwidth]{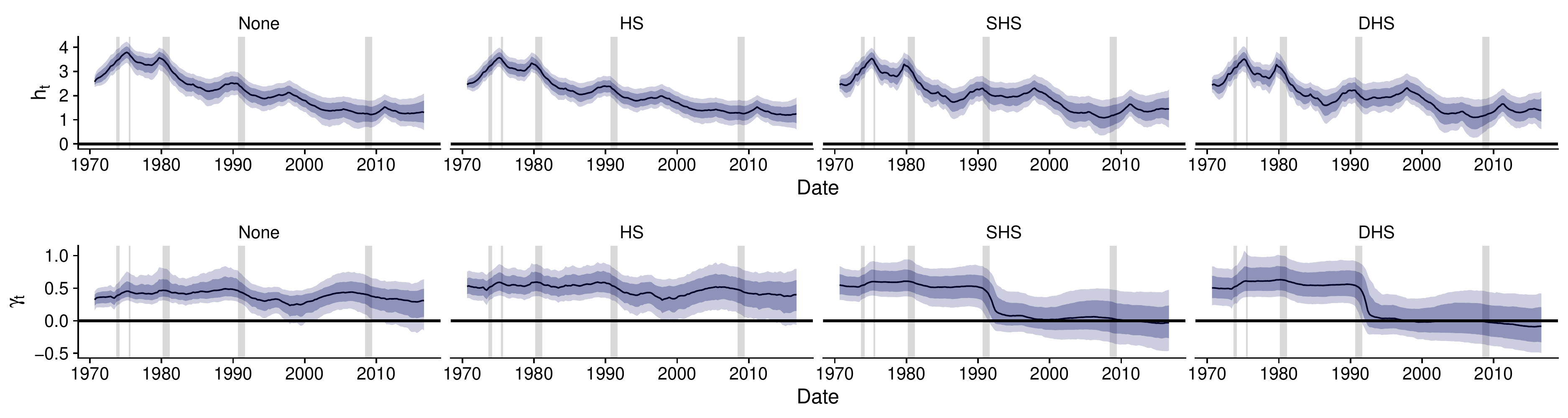}
\end{subfigure}
\begin{subfigure}[b]{\textwidth}
\vspace*{1em}\caption{Euro Area}
\includegraphics[width=\textwidth]{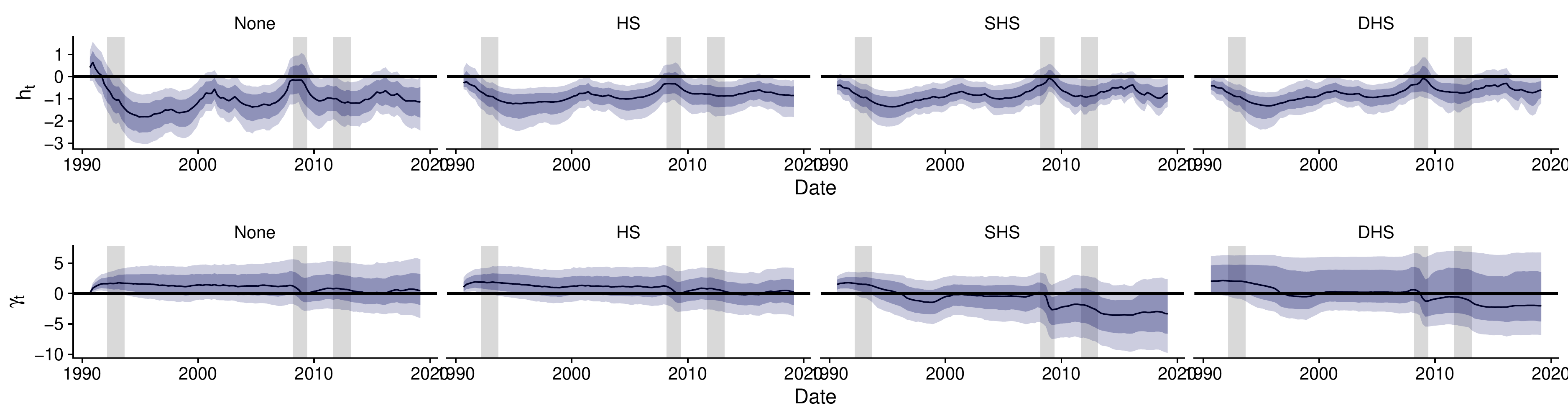}
\end{subfigure}
\end{center}
\caption{Time-varying volatilities $h_t$ and associated regression coefficients $\gamma_t$.}\label{fig:insample}\vspace*{-0.3cm}
\caption*{\footnotesize\textit{Note}: The black line is the posterior median estimate, alongside the $68$ percent (dark blue) and $90$ percent (light blue) posterior credible sets. Recessions are indicated as grey vertical bars.}
\end{figure}

Figure \ref{fig:insample} shows selected posterior credible intervals for the time-varying volatilities $h_t$ and the corresponding time-varying regression coefficients $\gamma_t$ over the full estimation period and across the three considered economies. For the US and the UK, the main impression is that the specific choice of the shrinkage specification plays a minor role for the estimates of $h_t$.  In the case of the EA, the specific choice of the prior seems to have some impact on the log-volatilities. In this case, any of the shrinkage priors appreciably reduces time-variation in $h_t$ for most periods except for the global financial crisis (GFC) in 2008/2009. Before and after that period, the error volatility process remains rather stable (as opposed to more rapidly changing log-volatilities in the no shrinkage case). 

Turning to the findings for $\gamma_t$ yields a different picture. While low frequency movements remain similar across shrinkage priors, some interesting differences arise. Shrinkage specifications that imply time-varying shrinkage (i.e. SHS and DHS) allow for sharp movements in $\gamma_t$ for selected periods and across economies. For instance, in the US we observe a pronounced change in the relationship between inflation and inflation volatility during the Volcker disinflation. A comparable appreciable decrease in $\gamma_t$ can also be observed in the UK during the crisis of the European Exchange Rate Mechanism (ERM) at the beginning of the 1990s. A similar decline, albeit more noisy, can be found during the GFC in the EA.

In sum (and with some exceptions) Figure \ref{fig:insample} shows that the original results of \cite{chan2017stochastic} remain remarkably robust with respect to different shrinkage priors. Exceptions arise especially during periods where the level of inflation experienced sharp changes (such as during the Volcker disinflation, the ERM and the GFC) and for EA data.

\subsection{Forecast results}
In this section, we analyze whether our set of shrinkage priors improves out-of-sample predictive performance within a real time forecasting exercise. We evaluate both point and density forecasts by means of root mean squared errors (RMSEs) and average log predictive likelihoods \citep[LPLs, see e.g.,][]{geweke2010comparing}. Each real time vintage is used to produce forecasts which are then evaluated using the final available vintage. 

We assess the merits of using shrinkage in the SVM model relative to the following competitors.
As in \citet{chan2017stochastic}, we use a random walk (RW) model as the benchmark for relative RMSEs and LPLs: $\pi_t = \pi_{t-1} + \eta_t$, with $\eta_t\sim\mathcal{N}(0,\sigma_{\eta}^2)$. Moreover, we include unobserved component models with stochastic volatility (UC-SV) as a special case of the UC-SVM model: $\pi_t = \tau_{t} + \epsilon_t$. We assume $\epsilon_t\sim\mathcal{N}(0,e^{h_t})$ with the state equation given by $h_t = \mu_h + \phi_h(h_{t-1} - \mu_h) + \nu_{t}$ and $\nu_t \sim \mathcal{N}(0, \sigma^2)$. UC-SV and UC-SVM are estimated using the four shrinkage priors (None, HS, SHS and DHS) discussed above. 
 
 Table \ref{tab:preds} presents forecasting results for different economies and shrinkage priors. In general (and with only very few exceptions) we find that all models improve upon the random walk. This holds true for both point and density forecasts, all economies and forecast horizons considered. Only in the case of density forecast accuracy for EA inflation we find the random walk to yield more precise predictions. The strong performance of the UC-SVM model \textit{without} shrinkage confirms the findings reported in \cite{chan2017stochastic}.
 
\begin{table*}[ht]
\begin{center}
\caption{Predictive inference relative to the benchmark model.}\label{tab:preds}\vspace*{-0.75em}
\begin{threeparttable}
\footnotesize
\begin{tabular*}{\textwidth}{@{\extracolsep{\fill}}ld{1.3}d{1.3}d{1.3}d{1.3}}
\toprule
  & \multicolumn{2}{c}{\textbf{RMSE}} & \multicolumn{2}{c}{\textbf{LPL}}\\
  \cmidrule(lr){2-3}\cmidrule(lr){4-5}
\textbf{Model} & \multicolumn{1}{c}{UC-SV} & \multicolumn{1}{c}{UC-SVM} & \multicolumn{1}{c}{UC-SV} & \multicolumn{1}{c}{UC-SVM} \\ 
\midrule
\textbf{United States} & \multicolumn{4}{c}{\textit{One-quarter ahead}} \\ 
  None & 0.677 & 0.726 & 0.374 & 0.499 \\ 
  HS & 0.679 & 0.777 & 0.474 & 0.477 \\ 
  SHS & 0.678 & 0.737 & 0.370 & 0.489 \\ 
  DHS & 0.680 & 0.724 & 0.369 & 0.488 \\ 
  \cmidrule(lr){2-5}
& \multicolumn{4}{c}{\textit{One-year ahead}}\\
  None & 0.742 & 0.757 & 0.467 & 0.584 \\ 
  HS & 0.740 & 0.788 & 0.539 & 0.570 \\ 
  SHS & 0.743 & 0.768 & 0.453 & 0.592 \\ 
  DHS & 0.739 & 0.764 & 0.448 & 0.572 \\ 
  \midrule
\textbf{United Kingdom} & \multicolumn{4}{c}{\textit{One-quarter ahead}} \\ 
None & 0.924 & 0.803 & 0.100 & 0.313 \\ 
  HS & 0.807 & 0.810 & 0.245 & 0.289 \\ 
  SHS & 0.818 & 0.805 & 0.122 & 0.305 \\ 
  DHS & 0.832 & 0.806 & 0.126 & 0.305 \\ 
  \cmidrule(lr){2-5}
& \multicolumn{4}{c}{\textit{One-year ahead}}\\
  None & 0.915 & 0.820 & 0.612 & 0.873 \\ 
  HS & 0.796 & 0.868 & 0.764 & 0.856 \\ 
  SHS & 0.803 & 0.865 & 0.636 & 0.849 \\ 
  DHS & 0.823 & 0.873 & 0.638 & 0.850 \\ 
  \midrule
\textbf{Euro area} & \multicolumn{4}{c}{\textit{One-quarter ahead}} \\ 
  None & 0.859 & 0.789 & -0.294 &  0.000 \\ 
  HS & 0.837 & 0.816 & 0.125 & 0.108 \\ 
  SHS & 0.865 & 0.821 & -0.291 &  0.061 \\ 
  DHS & 0.867 & 0.815 & -0.297 &  0.091 \\ 
  \cmidrule(lr){2-5}
& \multicolumn{4}{c}{\textit{One-year ahead}}\\
  None & 0.765 & 0.868 & -0.140 &  0.094 \\ 
  HS & 0.774 & 0.858 & 0.228 & 0.204 \\ 
  SHS & 0.764 & 0.885 & -0.140 &  0.175 \\ 
  DHS & 0.771 & 0.864 & -0.146 &  0.205 \\ 
\bottomrule
\end{tabular*}
\begin{tablenotes}[para,flushleft]
\scriptsize{\textit{Notes}: All measures are relative to the random walk benchmark. RMSEs are ratios (smaller numbers indicate superior performance), LPLs are differences (larger numbers indicate superior performance).}
\end{tablenotes}
\end{threeparttable}
\end{center}
\end{table*}
  
We now investigate whether using shrinkage further improves predictive accuracy. Considering both density and point forecasts, this question is difficult to answer. For some economies, horizons and specifications, shrinkage priors seem to improve both point and density forecasting performance while for other configurations, shrinkage seems to slightly hurt predictive accuracy. But these differences (both negative and positive) are often very small.  There exist some cases where we find more pronounced improvements. For instance, the UC-SV model with shrinkage performs appreciably better in predicting UK inflation at both horizons and by considering RMSEs and LPLs than the no-shrinkage counterpart. Another example that provides evidence that shrinkage improves forecasts can be found for EA inflation density forecasts. In this case, any shrinkage prior yields better forecasts than the no-shrinkage specification. 
 
 Considering differences between the different shrinkage priors provides no clear winner of our forecasting horse race. In most cases, predictions are similar to each other. If we were to choose a preferred prior our default recommendation would be the HS specification. This is because it performs well across the different configurations and for both model classes considered. Especially in the case of the EA, we find the HS setup to provide favorable point and density forecasts (especially for the UC-SV model).
 
The key take away from this discussion is that the benchmark model introduced in \cite{chan2017stochastic} seems to work very well for all considered economies. Using shrinkage helps in some cases but also leads to slightly inferior predictive performance in others. However, these decreases in forecast accuracy are never substantial. By contrast, we observe several cases where shrinkage improves forecasts. And these improvements are substantial. Hence, as a general rule we can suggest to combine the SVM model with shrinkage priors since the risk of obtaining markedly weaker forecasts appears to be low while the chances that forecasts can be improved substantially are much higher.

\section{Concluding remarks}
In this note we have successfully replicated the findings in \cite{chan2017stochastic} both in a narrow and wide sense. We have shown that using several different shrinkage techniques has the potential to improve forecasts. While these gains are small on average, several cases emerge where improvements are more pronounced. More importantly, we never find situations where using shrinkage strongly decreases forecast performance.

\small{\setstretch{0.85}
\addcontentsline{toc}{section}{References}
\bibliographystyle{custom.bst}
\bibliography{lit}}

\end{document}